\documentclass[prx,reprint,superscriptaddress,twocolumn,showkeys]{revtex4-1}
\usepackage[utf8]{inputenc}
\usepackage{amsmath}
\usepackage{braket}
\usepackage{graphicx}
\usepackage{tabularx}
\usepackage{rotating}
\usepackage{pgfplots}
\pgfplotsset{compat=1.15}
\usepackage{csquotes}
\usepackage{hhline}
\usepackage{amssymb}

\usepackage{dcolumn}
\usepackage[colorlinks=true,linkcolor=blue,citecolor=blue,urlcolor=blue]{hyperref}
\usepackage{braket}
\usepackage{float}
\usepackage{listings}
\usepackage{color}
\usepackage{subcaption}
\usepackage{placeins}
\definecolor{mygreen}{rgb}{0,0.6,0}
\definecolor{mygray}{rgb}{0.5,0.5,0.5}
\definecolor{mymauve}{rgb}{0.58,0,0.82}

\newcolumntype{C}{>{\centering\arraybackslash}X}

\begin{document}

\title{Factorization of large tetra and penta prime numbers on IBM quantum processor}

\author{Ritu Dhaulakhandi}
\email{ritudhaulakhandi3626@gmail.com}
\affiliation{Department of Physics,\\ Indian Institute of Science Education and Research, Pune, 411008, Maharastra, India}
\author{Bikash K. Behera}
\email{bikas.riki@gmail.com}
\affiliation{Bikash's Quantum (OPC) Private Limited, Balindi, Mohanpur 741246, Nadia, West Bengal, India}
\author{Felix J. Seo}
\email{jaetae.seo@hamptonu.edu}
\affiliation{Department of Physics,\\ Hampton University, Hampton, Virginia, 23668, USA}

\begin{abstract}
The factorization of a large digit integer in polynomial time is a challenging computational task to decipher. The exponential growth of computation can be alleviated if the factorization problem is changed to an optimization problem with the quantum computation process with the generalized Grover's algorithm and a suitable analytic algebra. In this article, the generalized Grover's protocol is used to amplify the amplitude of the required states and, in turn, help in the execution of the quantum factorization of tetra and penta primes as a proof of concept for distinct integers, including 875, 1269636549803, and 4375 using 3 and 4 qubits of IBMQ Perth (7-qubit processor). The fidelity of quantum factorization with the IBMQ Perth qubits was near unity.
\end{abstract}

\maketitle

\section{Introduction}

Computation problems are often classified by the difficulty of getting their solutions. The concept of difficulty is explained by the theory of computational complexity \cite{qfa_DingZhuJWS2011}, which specifies the time required to perform a problem using a specific approach. It is well known that cryptography technology utilizes the difficulty of factorizing large numbers to secure data storage and information transmission \cite{qfa_Traversaa2017}. If the factorization is a polynomial-time problem, the security system is not secure. In order to factorize an n-bit integer on a quantum computer, Shor presented a polynomial-time approach in 1994 \cite{qfa_ShorSIAM1997}. Later, the procedure is put into practice by factorizing the numbers $N=15$ and $N=21$ \cite{qfa_VandersypenNature2001} and $N=15$ \cite{qfa_LopezNatPhot2012}. The only shortcoming of implementing Shor's algorithm is the need for robust error correction schemes and noise-free qubits \cite{qfa_VandersypenNature2001, qfa_LopezNatPhot2012, qfa_SmolinNature2013}. More specifically, a high number of qubits are needed to factorize an integer without knowing the solution beforehand. For instance, using Shor's algorithm without knowing the solution, factorizing the number $15$ would require at least $8$ qubits (and more for error correction) \cite{qfa_DattaniarXiv2014}.

Many alternative methods \cite{qfa_Yan2015} have been developed to overcome the disadvantage of implementing Shor's algorithm. The adiabatic quantum computation \cite{qfa_PengPRL2008} method, which transforms the factorization problem into an optimization problem \cite{qfa_BurgesMicrosoft2002}, is one of these techniques. The factorization target number's binary system multiplication table is expressed in variable form. Using the reduction technique, the problem is reduced to a list of equations \cite{qfa_DattaniarXiv2014}. A complex Hamiltonian is expressed using those equations. The ground states of the Hamiltonian carry the solutions (zero eigenvalue states). The use of quantum annealing methods and computational algebraic geometry to factorize the bi-primes was also suggested by Dridi and Alghassi in their 2017 paper \cite{qfa_DridiScirep2017}. The theoretical and experimental quantum factorization have been reported utilizing Shor's algorithm \cite{qfa_VandersypenNature2001,qfa_LopezNatPhot2012,qfa_BocharovPRA2017}, adiabatic quantum computation \cite{qfa_PengPRL2008,qfa_WangFP2022,qfa_DattaniarXiv2014,qfa_XuPRL2012,qfa_LiarXiv2017,qfa_XuPRL2017}, and quantum annealing principles \cite{qfa_DridiScirep2017}. The final number of variables in the equation, calculated using the minimization strategy for a certain integer, determines the total amount of qubits required for the experimental quantum factorization. The largest integers factorized using different algorithms are listed in Table \ref{qfa_table1}.

\begin{table}
\centering
\caption{\emph{Quantum Factorization Methods}}
\begin{tabular}{ccc}
\hline
\hline
Largest number & Protocol used & No. of qubits \\ \hline
$21$ \cite{qfa_LopezNatPhot2012}   & Shor's algorithm         & $10$ \\
$21$ \cite{qfa_PengPRL2008}   & Quantum adiabatic         & $3$ \\
   & algorithm         & \\
$1829$ \cite{qfa_SelvarajanSR2021}   & Quantum variational & $9$ \\
& imaginary time         & \\
& evolution         & \\
$1005973$ \cite{qfa_PengSCPMASR2019}   & Quantum Annealing         & $89$ \\
$4088459$ \cite{qfa_DasharXiv2018}   & Minimization         & $2$ \\ 
\hline
\hline
\end{tabular}
\label{qfa_table1}
\end{table}

This article utilizes the minimization method \cite{qfa_XuPRL2012,qfa_DasharXiv2018} for pre-processing, like the adiabatic approach \cite{qfa_WangFP2022}. After applying the minimization method, the Hamiltonian is written down using the final equations. A unitary operator is then defined as the exponential function of the said Hamiltonian. The unitary operator marks the Hamiltonian's ground states (states with zero eigenvalues). The states obtained after the application of the Hamiltonian unitary operator will be referred to as marked states in this article. To distinguish the marked states from the unmarked states, the generalized Grover's method \cite{qfa_LiuIJTP2014,qfa_liPLA2007} was used to amplify the marked states. The initial state (uniform superposition of qubit system) progresses to the target state (marked states) by repeated application of the oracle and diffuser operator. The multiple target states' amplitudes are increased by the generalized Grover's technique. The quantum factorization protocol was utilized for the quantum computation experiment to factorize the integers, including $875$ and $1269636549803$ using $3$-qubit systems and $4375$ using $4$-qubit system on IBM's quantum processor ibmq\_perth ($7$-qubit) for the proof-of-concept. The following sections describe the quantum factorization protocol and the properties of the factors of integers used in detail.

\section{Background}
\subsection{RSA and Quantum factorization}

When factoring an integer, the needed time order is $O(bk)$ with the $k$-th order of $b$-bit number, indicating that the factorization takes a polynomial amount of time. Equally challenging as the factorization problem is figuring out how many prime factors there are in an integer. There are no effective number-theoretic functions to determine the number of prime factors of an integer in number theory \cite{qfa_ShoupCUP2008}. The sieve theory, for instance, calculates the approximate number of prime factors in an integer. The inability of sieve theory to distinguish between numbers having an odd or even number of prime elements is known as the parity problem. A lot of cryptographic protocols benefit from the factorization problem's complexity. Often employed in cryptography, the RSA \cite{qfa_RivestCACM1978} relies on the challenge of factoring a big bi-prime integer in polynomial time to secure data transmission. With the RSA protocol, a public key is made available based on a massive bi-prime number. The decryption key, or private key, is different from the public key \cite{qfa_DiffieIEEE1976}. The two prime elements of the employed bi-prime number are kept a secret. Huge bi-prime numbers are used to make the RSA encryption unbreakable. The steps involved in RSA encryption and decryption \cite{qfa_SihotangJPCS2020,qfa_AnadaJISA2019} are listed below.

\begin{itemize}
    \item Choose two unique prime numbers. Calculate their product ($n$).
    \item Carmichael's totient function ($\lambda(n)$) is evaluated using the least common multiple to fix the range of $e$ between $1$ and $\lambda(n)$.
    \item Using $e$ and modular multiplicative inverse, the encryption and decryption functions are defined for the public and private keys, respectively.
\end{itemize}

Shor's approach, which is implemented on a quantum computer, can be used to find an integer's prime factors \cite{qfa_ShorSIAM1997}. Shor demonstrates how RSA encryption may be broken using a quantum computer approach to factor huge integers in polynomial time. Based on the complexity of the factorization issue, it enables a quantum computer to decrypt the public key. This demands new cryptographies that provide security against the capability of quantum algorithms \cite{qfa_LuyJISA2016}. Therefore, the quantum cryptography field has been intensively studied for secure data transmission \cite{qfa_GisinRMP2002,qfa_BhattJEST2019}.

\subsection{Existing Methods}

Burges introduced the quantum adiabatic theorem in 2001 \cite{qfa_BurgesMicrosoft2002}. In the quantum adiabatic method, a Hamiltonian is constructed from the multiplication table of the integer to be factorized. By reducing the number of variables in the problem, this strategy is effective for integers with unique features \cite{qfa_DattaniarXiv2014,qfa_XuPRL2012}. Because the minimizing method makes the factorization model more complex, it cannot be used in all cases. This prompted a search for a more inclusive prime factorization procedure. On the basis of the quantum annealing principle, certain generalized models have been put out; however, they are still constrained by the hardware capabilities of quantum machines \cite{qfa_DridiScirep2017,qfa_JiangSR2018}. A new prime factorization model that uses less quantum annealing and fewer qubits was proposed by Wang, B. \emph{et al.} in 2020.

\begin{figure}
\centering
\includegraphics[width=\linewidth]{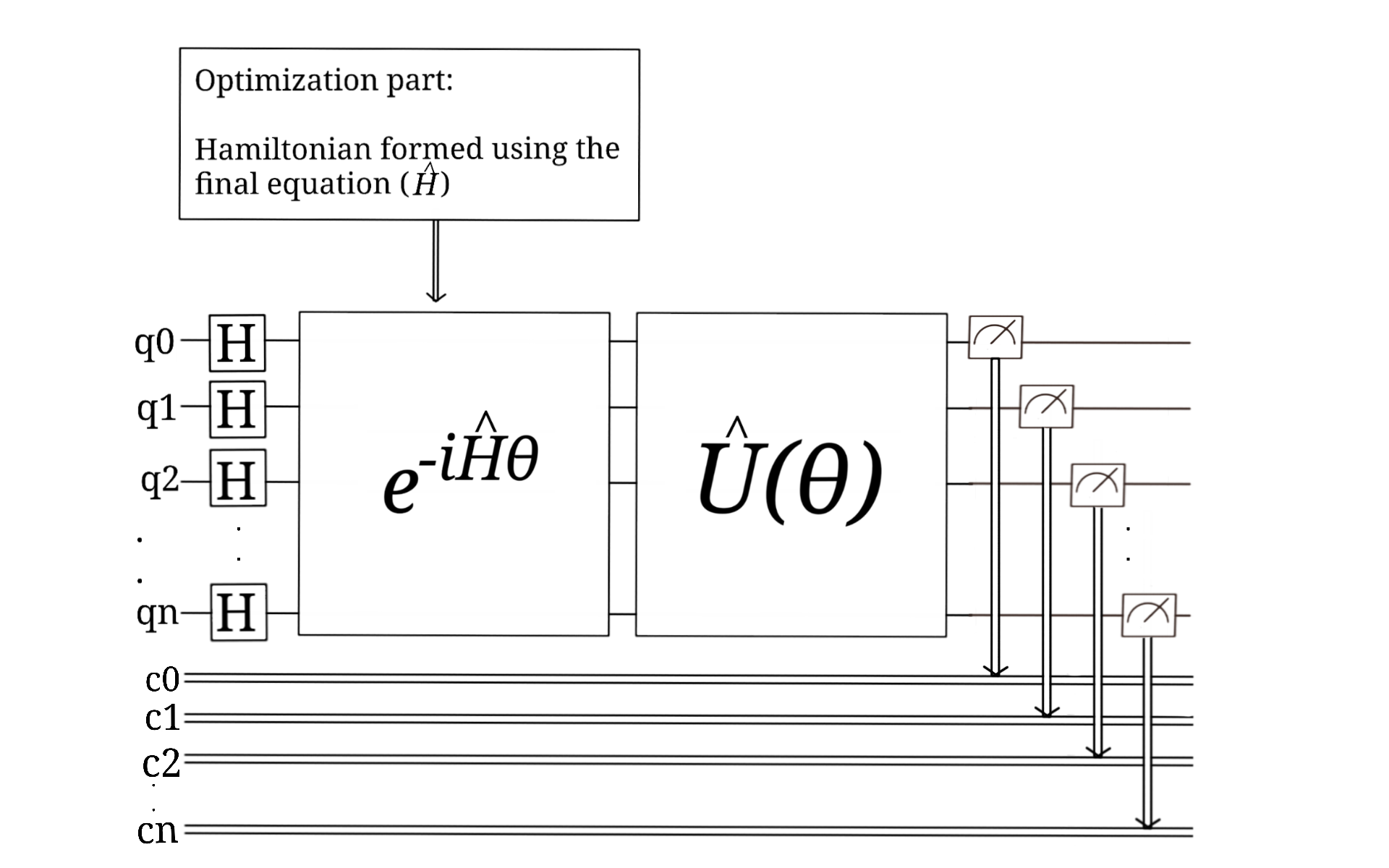} 
\caption{\textbf{Schematic diagram of Factorization Protocol} The protocol begins with the optimization part. The integer's prime factors are expressed in binary form, and their product is evaluated to form equations. Using the minimization strategy, the equations' variable count is decreased. The final set of equations obtained is used to write down the Hamiltonian ($\hat{H}$). The operator is defined by this Hamiltonian. The superposition state is initially produced on the qubits using the Hadamard gate. Then, the operator described using the Hamiltonian marks the Hamiltonian's ground states (states with eigenvalue zero) by applying a phase shift to the states. To boost the amplitude of the marked states, the operator $\hat{U}(\theta)$ is defined using the generalized Grover's search technique. Finally, the measurements are recorded for the different computational basis to obtain the experimental density matrix.}
\label{qfa_QFS}
\end{figure}

This article implements the quantum factorization protocol for factorizing tetra and penta prime numbers. The protocol includes a pre-processing part and a quantum computation part and overcomes the shortcoming of Shor's algorithm. The pre-processing part changes the factorization problem into an optimization problem. The prime factors of the integer are expressed in binary form. Then the binary product of the prime factors is evaluated. A set of equations are obtained from the binary product. The number of variables present in these equations is reduced using binary arithmetic rules. This simplification is called minimization \cite{qfa_WangFP2022,qfa_XuPRL2012,qfa_DasharXiv2018}. After obtaining the final set of equations, the bit variables are mapped with the quantum operators. This mapping of variables to operators is defined so that the Hamiltonian encodes the solutions (required bit values) as its ground states (states with eigenvalue as zero). The zero eigenvalue states of the Hamiltonian are then marked by a conditional phase shift carried out by the unitary operator defined using the Hamiltonian. The marked states become more pronounced using the generalized Grover's algorithm. The factorization problem is shown schematically in Fig. \ref{qfa_QFS}. A few integers are factorized as a confirmation for the quantum computation experiment. The tetra prime integers $875$ and $1269636549803$ consist of four prime factors. The penta prime integer $4375$ consists of five prime factors. The experimental results' fidelity \cite{qfa_Ansari2020} was examined with the help of quantum state tomography \cite{qfa_Altepeter2004}. The following sections include the quantum computation experiment of prime factorization based on the protocol.

\section{Methodology}

The following general procedure is applied for the factorization problem. Let $N$ denote an odd composite integer with $\alpha$ number of prime factors. Each prime factor of $N$ is represented as $n_i$, where $i$=$1$, $2$, . . ., $\alpha$ ($\alpha\in\mathbb{N}$). The prime factors are written in binary form, and the binary product is evaluated in the form of variables. Then the set of equations are given by the binary product $(n_1)_{bin} \times (n_2)_{bin}….(n_{\alpha})_{bin}=N_{bin}$ where $(n_i)_{bin}$ indicates integer $n_i$ in binary system (or representation as a binary number) \cite{qfa_BurgesMicrosoft2002,qfa_XuPRL2012,qfa_WangFP2022,qfa_DattaniarXiv2014,qfa_LiarXiv2017}. The equations are optimized by reducing the number of variables with the arithmetic of binary numbers, called the minimization method. The prime factors $n_{\alpha}$  of the $N$ satisfy the property that all the prime factors have an equal number of digits in their binary form (number of digits in $(n_{\alpha})_{bin}$ is same for all $\alpha$). And

\begin{equation}
    (n_{1})_{bin}=(n_{2})_{bin}=....=(n_{\alpha-1})_{bin}\not=(n_{\alpha})_{bin}
    \label{equn1}
\end{equation}

The case for $\alpha=2$ is worked out to understand the optimization part more clearly. Let $N$=$35$ and $n_1$ and $n_2$ be the prime factors of $N$. The number of digits in $(n_1)_{bin}$=$(n_2)_{bin}$=$3$, and $(n_1)_{bin}\equiv(1b_11)$, $(n_2)_{bin}\equiv(1c_11)$.

\begin{figure}
\centering
\includegraphics[width=0.9\linewidth]{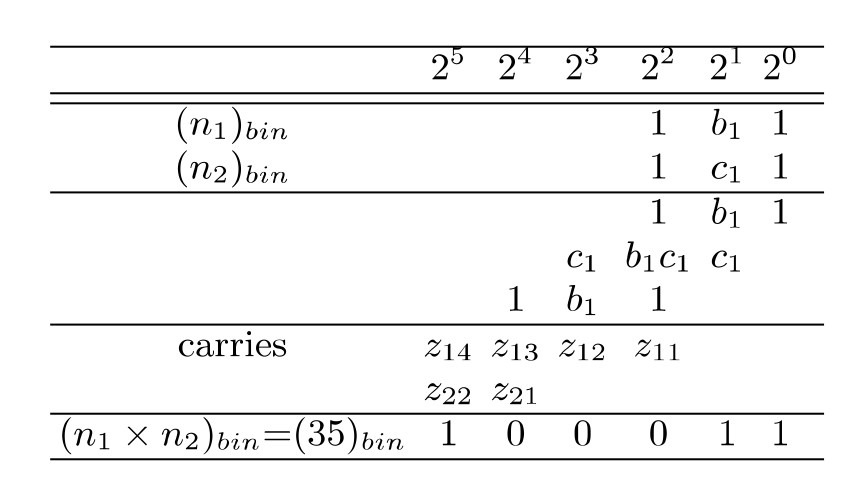} 
\caption{\emph{Multiplication table for $\alpha$=$2$ case, $N$=$35$.}}
\label{table0}
\end{figure}

From Fig. \ref{table0}, the equations obtained after adding each column are as follows:

\begin{eqnarray}\label{meq1}
    b_1+c_1=1+2z_{11}\\ \nonumber
    2+b_1c_1+z_{11}=0+2z_{12}+4z_{21}\\ \nonumber
    b_1+c_1+z_{12}=0+2z_{13}+4z_{22}\\ \nonumber
    1+z_{13}+z_{21}=0+2z_{14}\\ \nonumber
    z_{14}+z_{22}=1
\end{eqnarray}

These equations are further simplified with binary number arithmetic to obtain:

\begin{eqnarray}\label{m2}
    z_{11}=0\\ \nonumber
    z_{12}=1\\ \nonumber
    z_{21}=0\\ \nonumber
    z_{13}=1\\ \nonumber
    z_{22}=0\\ \nonumber
    z_{14}=1
\end{eqnarray}

Therefore, the equation used to formalize the Hamiltonian for $N$=$35$ is an equation in one variable, $q_1$. The minimization method reduced the $N$=$35$ factorization problem to a single variable problem.

\begin{eqnarray}\label{m3}
    b_1+c_1=1\\ \nonumber
    b_1c_1=0\\ \nonumber
    \implies c_1-c_1^2=0
\end{eqnarray}

The set of equations obtained after evaluating the binary product ($(n_1)_{bin}\times(n_2)_{bin}$) in Eq. \ref{meq1} have $8$ variables. The minimization method reduces the number of variables, as seen in Eq. \ref{m3}. The minimization method uses arithmetic of binary numbers (Eq. \ref{m2}) where each binary digit takes the value of either $0$ or $1$. The final equation is used to write down the Hamiltonian. In this article, the quantum factorization procedure is carried out for the odd composite number with $\alpha=4$, and $5$ that satisfies the mentioned property (Eq. \ref{equn1} and the number of digits in binary form are the same for all prime factors). The details for the quantum computation part and the results of tetra and penta prime quantum factorization are explained.

\section{Results}
\subsection{Quantum factorization of tetra prime number}

The binary representations of the prime factors $p$, $q$, $r$, and $s$ of $N$=$875$ are ($1p_11$), ($1q_11$), ($1r_11$), and ($1s_11$), respectively.

\begin{figure*}
\centering
\includegraphics[width=0.6\linewidth]{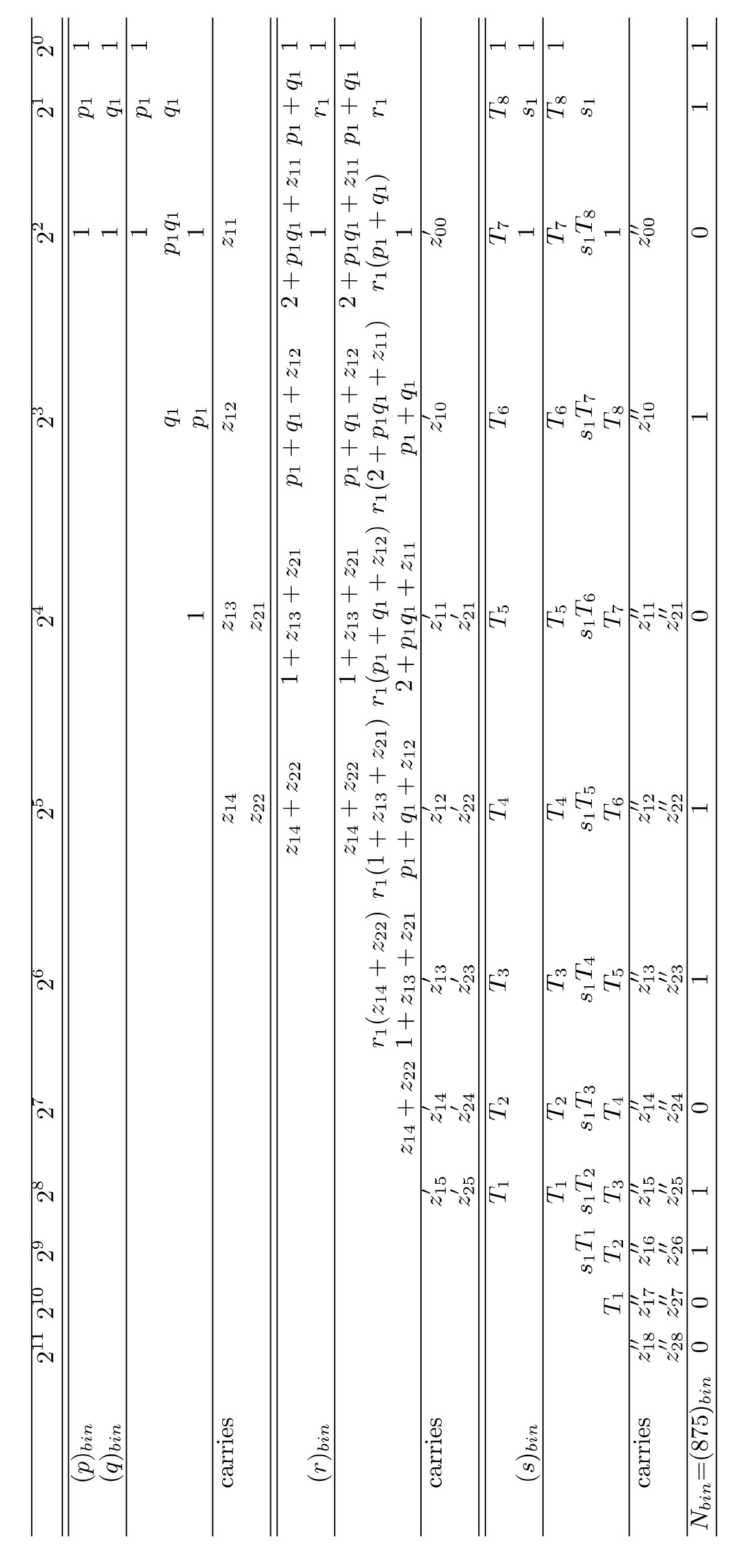} 
\caption{\emph{Multiplication table for $\alpha$=$4$ case, $N$=$875$.\\
$T_1$=$z_{15}'+z_{25}'$, $T_2$=$z_{14}+z_{22}+z_{14}'+z_{24}'$, $T_3$=$r_1(z_{14}+z_{22})+1+z_{13}+z_{21}+z_{13}'+z_{23}'$, $T_4$=$z_{14}+z_{22}+r_1(1+z_{13}+z_{21})+p_1+q_1+z_{12}+z_{12}'+z_{22}'$, $T_5$=$1+z_{13}+z_{21}+r_1(p_1+q_1+z_{12})+2+p_1q_1+z_{11}$, $T_6$=$p_1+q_1+z_{12}+r_1(2+p_1q_1+z_{11})+p_1+q_1$, $T_7$=$3+p_1q_1+z_{11}+r_1(p_1+q_1)$, $T_8$=$p_1+q_1+r_1$}}
\label{t1}
\end{figure*}

\begin{figure*}
\centering
\includegraphics[width=\linewidth]{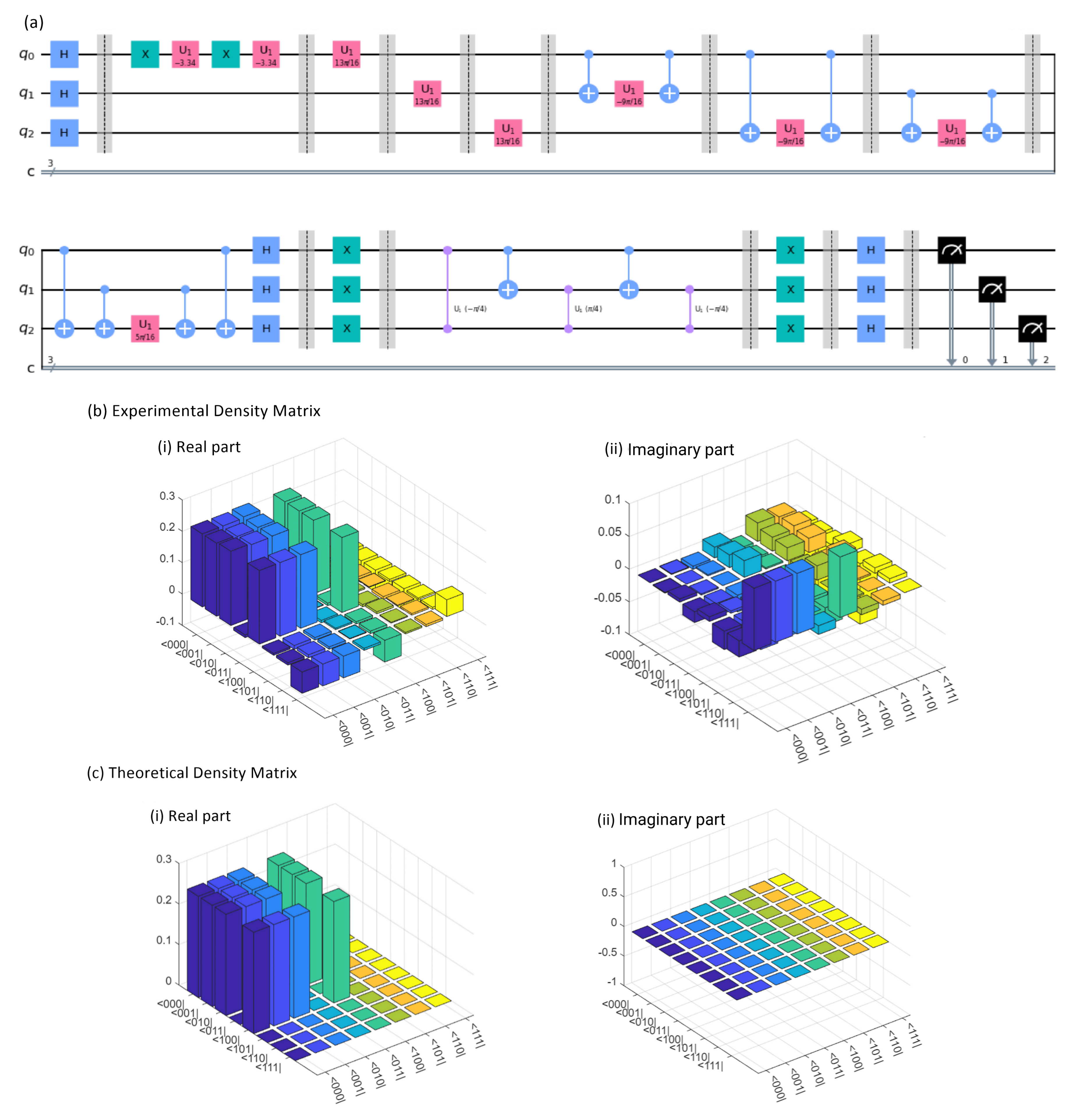} 
\caption{\textbf{(a)Quantum factorization circuit, (b) Experimental density matrix (EDM), and (c) Theoretical density matrix (TDM)} The circuit, which uses $3$ qubits to factorize $N$=$875$, is implemented on IBM's quantum processor. The experiment and simulation are performed on the ibmq$\_$perth system, a $7$-qubit processor. The final measurements are obtained in the $Z$-basis, forming the EDM. The EDM has a non-zero complex part, as shown in (b)(ii). As indicated in (b) and (c), the numerical values of the real component and imaginary component of the density matrix elements are shown separately. The fidelity of the result was found to be 0.9683 (The EDM and fidelity calculation are provided in the link \cite{qfa_github}). The $U_1$ gate is a $2\times2$ diagonal matrix with diagonal entries $\left(1,e^{i\phi}\right)$ which is equivalent to the operation of $e^{\frac{i\phi}{2}}R_z(\phi)$ where $R_z$ performs a rotation around z-axis by an amount $\phi$.}
\label{qfa_Fig1}
\end{figure*}

The set of equations found from the binary product (Fig. \ref{t1}) followed by minimization are:

\begin{eqnarray}\label{qfa_eq_1}
    && p_1 + q_1 + r_1 + s_1 = 1\nonumber \\
    && p_1q_1 + q_1r_1 + p_1r_1 + s_1p_1 + s_1q_1 + s_1r_1 = 0 \nonumber \\
    && p_1q_1r_1 + p_1r_1s_1 + p_1q_1s_1 + q_1r_1s_1 = 0
\end{eqnarray}

which are reduced to a single equation with three variables:

\begin{eqnarray}\label{qfa_eq_2}
    && -p_1-q_1-r_1+2p_1^2+2q_1^2+2r_1^2-p_1^3-q_1^3-r_1^3\nonumber \\ 
    &&+4p_1r_1+4p_1q_1+4q_1r_1-3p_1^2q_1-3p_1^2r_1-3q_1^2p_1\nonumber \\ 
    &&-3q_1^2r_1-3r_1^2p_1-3r_1^2q_1-5p_1q_1r_1 = 0
\end{eqnarray}

After multiplying Eq. \ref{qfa_eq_2} with $-1$ on both sides, the variables $(p_1,q_1,r_1)$ are mapped with the operators $(\hat{a_1},\hat{a_2},\hat{a_3})$ where $\hat{a_i}=\frac{I-\sigma_z^i}{2}$, $I$ is the $1$-qubit identity operator, and the $\sigma_z^i$ operator in the quantum circuit acting on the $i$th qubit is the Pauli $Z$ operator. This mapping transforms the Hamiltonian from a variable equation to a diagonal matrix form with non-negative number entries. The ground states of Hamiltonian encode the solutions. For example, if $H\ket{qu_1}\ket{qu_2}\ket{qu_3}$=$0\ket{qu_1}\ket{qu_2}\ket{qu_3}$ then $b_1$=$p_1$, $b_2$=$q_1$, and $b_3$=$r_1$ are the required bit values. For example, from Eq. \ref{H1} it is observed that the Hamiltonian for $N$=$875$ has ground states (states with zero eigenvalue) $\ket{000}$, $\ket{001}$, $\ket{010}$, and $\ket{100}$. Operator $\hat{a_i}$ also satisfies the property $\hat{a_i}^2$=$\hat{a_i}$. As a result, the following is the Hamiltonian for the factorization problem:

 \begin{equation}\label{qfa_eq_3}
        \hat{H} = 5\hat{a_1}\hat{a_2}\hat{a_3}+2\hat{a_1}\hat{a_2}+2\hat{a_1}\hat{a_3}+2\hat{a_2}\hat{a_3}
 \end{equation}
 
 The $\hat{a_i}$ operators are substituted in the above Hamiltonian to obtain:
 
 \begin{eqnarray}\label{qfa_eq_4}
      \hat{H}=&&\frac{17}{8}(I_{3})-\frac{13}{8}(\sigma_z^1\otimes I\otimes I+I\otimes\sigma_z^2\otimes I+I\otimes I\otimes \sigma_z^3)\nonumber\\
      &&+\frac{9}{8}(\sigma_z^1\otimes\sigma_z^2\otimes I+\sigma_z^1\otimes I\otimes \sigma_z^3+I\otimes\sigma_z^2\otimes \sigma_z^3)\nonumber\\
      &&-\frac{5}{8}(\sigma_z^1\otimes \sigma_z^2\otimes \sigma_z^3)
 \end{eqnarray}
 
\begin{equation}\label{H1}
    \hat{H}= \left[ {\begin{array}{cccccccc}
        0 & 0 & 0 & 0 & 0 & 0 & 0 & 0  \\
        0 & 0 & 0 & 0 & 0 & 0 & 0 & 0  \\
        0 & 0 & 0 & 0 & 0 & 0 & 0 & 0  \\
        0 & 0 & 0 & 2 & 0 & 0 & 0 & 0  \\
        0 & 0 & 0 & 0 & 0 & 0 & 0 & 0  \\
        0 & 0 & 0 & 0 & 0 & 2 & 0 & 0  \\
        0 & 0 & 0 & 0 & 0 & 0 & 2 & 0  \\
        0 & 0 & 0 & 0 & 0 & 0 & 0 & 11  \\
    \end{array} }\right]
\end{equation}
 
\begin{table}
\centering
\caption{\emph{For $N$=$875$. The phase shift relative to the state $\ket{000}$ when the operator $e^{-i\hat{H}\theta}$ is applied to the $z$-basis states.}}
\begin{tabular}{cccc}
\hline
\hline
Quantum  & Relative & Quantum & Relative \\
State  & Phase shift & State & Phase shift \\ \hline
$\ket{000}$   & $0$         & $\ket{100}$   & $0$   \\
$\ket{001}$   & $0$         & $\ket{101}$   & $-2\theta$   \\ 
$\ket{010}$   & $0$         & $\ket{110}$   & $-2\theta$   \\
$\ket{011}$   & $-2\theta$         & $\ket{111}$   & $-11\theta$   \\
\hline
\hline
\end{tabular}
\label{qfa_table2}
\end{table}
 
 where $I_3$ is the $3$-qubit identity operator. The Hamiltonian's ground state ($\ket{qu_1qu_2qu_3}$) are the solutions for the binary digits ($p_1$=$qu_1$, $q_1$=$qu_2$, and $r_1$=$qu_3$). Since $\hat{H}$ is a diagonal matrix, the operator $e^{-i\hat{H}\theta}$ can be expressed as a diagonal matrix too. The $e^{-i\hat{H}\theta}$ operator applies a conditional phase shift (relative to $\ket{000}$ state) to the initial superposition state as shown in Table \ref{qfa_table2}. The initial uniform superposition state $\ket{\psi_0}$=$\frac{1}{2\sqrt{2}}\Sigma^{i=7}_{i=0} \ket{i}$ is obtained by applying Hadamard gate $H^{\otimes3}$ to the $\ket{000}$ state. The ground states of Hamiltonian are called marked states after applying the $e^{-i\hat{H}\theta}$ operator. For the purpose of differentiating the marked states from the unmarked ones, the amplitudes of the marked states are enhanced. The amplification procedure is carried out by an exact search method. This is accomplished by using an oracle $\hat{U}(\theta)$ that was obtained using the generalized Grover's search technique \cite{qfa_LiuIJTP2014}. The way the search algorithm operates is similar to how resonance works. By re-expressing the state $\ket{\psi_0}$, one may determine the phase shift angle ($\theta$). $\ket{\psi_0}$ can be written in terms of the normalized sum of marked states $\ket{x_0}$ ($\ket{x_0}$=$\frac{1}{\sqrt{M}}\sum_{t=0}^{M-1}\ket{MS_t}$, where $\ket{MS_t}$ are the marked states) and the normalized sum of remaining states $\ket{x^{\perp}_0}$ ($\ket{x^{\perp}_0}$=$\frac{1}{\sqrt{N-M}}\sum_{f=0}^{N-M-1}\ket{S_f}$, where $\ket{S_f}$ are the remaining states). The expression of $\ket{\psi_0}$ in terms of $\ket{x_0}$ and $\ket{x^{\perp}_0}$ is:
 
 \begin{equation}
     \ket{\psi_0}=\sin{\phi}\ket{x_0}+\cos{\phi}\ket{x^{\perp}_0}
 \end{equation}
 
where the angle $\phi$ is the reflection angle of the uniform superposition state w.r.t. the unmarked states. The phase shift angle $\theta$ and $\phi$ are related by the following equation: $\theta$=$2\sin^{-1}{(\frac{\sin{\frac{\pi}{4j+2}}}{\sin{\phi}})}$, where $j$ is the smallest number of iterations of Grover's protocol necessary to maximize the amplitude of the solution states while minimizing the amplitude of the remaining states \cite{qfa_LiuIJTP2014,qfa_liPLA2007}. Quantum tomography is used to examine the accuracy of the experimental result after the quantum circuit results have been obtained. Quantum tomography uses a sequence of measurements on several bases to capture the complete quantum state. It provides fidelity between the density matrices of experimental data and theoretical values for the factorization problem. The TDM is given as $\rho^T$=$\ket{\Psi}\bra{\Psi}$, where $\ket{\Psi}$=$\hat{U}(\theta)e^{-i\hat{H}\theta}\ket{\psi_0}$ is the final state obtained after implementing the quantum circuit for a given factorization problem. The EDM is given by the Stokes parameters $S_a$ \cite{qfa_Altepeter2004, qfa_JamesPRA2001} and Pauli matrices $\sigma_a$. The EDM is given as $\rho_{3}^E$=$\frac{1}{N}\Sigma_{i,j,k}(S_i \otimes S_j \otimes S_k)(\sigma_i \otimes \sigma_j \otimes \sigma_k)$ for the $3$ qubit system, and $\rho_{4}^E$=$\frac{1}{N}\Sigma_{i,j,k,l}(S_i \otimes S_j \otimes S_k \otimes S_l)(\sigma_i \otimes \sigma_j \otimes \sigma_k \otimes \sigma_l)$ for the $4$ qubit system, where $i$, $j$, $k$, $l$ go from zero to three, and $\sigma_a$ belongs to the set of $\{I$,$\sigma_X$,$\sigma_Y$,$\sigma_Z\}$. The following are the Stokes parameters:
 \begin{eqnarray}
     &&S_3=P_{\ket{0_z}}-P_{\ket{1_z}}\nonumber\\ \nonumber\\
     &&S_2=P_{\ket{0_y}}-P_{\ket{1_y}}\nonumber\\ \nonumber\\
     &&S_1=P_{\ket{0_x}}-P_{\ket{1_x}}\nonumber\\ \nonumber\\
     &&S_0=1
 \end{eqnarray}

where the probability of discovering state $\ket{i}$ in basis $j$ is $P_{\ket{i_j}}$. For Pauli-$Z$ matrix ($\sigma_z$), the eigenvectors are $\ket{0}$ and $\ket{1}$. In the Bloch sphere, $Z$ basis measurement gives the probability of states $\ket{0}$ and $\ket{1}$, $X$ basis measurement gives the probability of finding states $\ket{+}$=$H\ket{0}$ and $\ket{-}$=$H\ket{1}$, where H is the Hadamard matrix, and $Y$ basis measurement gives the probability of finding states $\ket{+i}$=$SH\ket{0}$ and $\ket{-i}$=$SH\ket{1}$, where $SS^{\dag}$=$\mathbb{I}$ \cite{qfa_Microsoft}.
 
\begin{eqnarray}\label{qfa_eq_7}
H=\frac{1}{\sqrt{2}}\left[{\begin{array}{cc}
          1&1  \\
          1&-1 
     \end{array}}\right]\nonumber
     \hspace{0.5cm}S=\left[{\begin{array}{cc}
          1&0  \\
          0&i 
     \end{array}}\right]\nonumber\\
     X=\left[{\begin{array}{cc}
          0&1  \\
          1&0 
     \end{array}}\right]   \nonumber
     \hspace{0.5cm}Y=\left[{\begin{array}{cc}
          0&-i \\
          i&0 
     \end{array}}\right]\nonumber\\
     Z=\left[{\begin{array}{cc}
          1&0  \\
          0&-1 
     \end{array}}\right]
     \hspace{0.5cm}\mathbb{I}=\left[{\begin{array}{cc}
          1&0  \\
          0&1 
     \end{array}}\right]
\end{eqnarray}
 
 The fidelity value is calculated using the $\rho^T$ and $\rho^E$ matrices which tells the extent of overlap between EDM and TDM, $F(\rho^T,\rho^E)$=$Tr(\sqrt{\sqrt{\rho^T}\rho^E\sqrt{\rho^T}})$.
 
 For $N$=$875$, the four solutions obtained ($|p_1q_1r_1\rangle$) are as expected. Four marked states are obtained from the Hamiltonian's ground states by applying $e^{-i\hat{H}\theta}$. The exact search algorithm searches for the four marked solution states and amplifies their amplitude. The phase shift angle to achieve maximum amplification is given as $\theta=2sin^{-1}\left(\frac{\sin{\frac{\pi}{4j+2}}}{\sin{\phi}}\right)$ where $\phi$=$\frac{\pi}{4}$ and $j$=$1$ iteration \cite{qfa_LiuIJTP2014,qfa_liPLA2007}. The EDM is given in Fig. \ref{qfa_Fig1} (b). The TDM for $N$=$875$ is given as $\rho^T$=$\ket{\Psi}\bra{\Psi}$ where $\ket{\Psi}=\frac{1}{2}\left[\ket{000}+\ket{001}+\ket{010}+\ket{100}\right]$.
\begin{equation}\label{qfa_eq_8}
    \rho^{T}= \frac{1}{4} \left[ {\begin{array}{cccccccc}
        1 & 1 & 1 & 0 & 1 & 0 & 0 & 0  \\
        1 & 1 & 1 & 0 & 1 & 0 & 0 & 0  \\
        1 & 1 & 1 & 0 & 1 & 0 & 0 & 0  \\
        0 & 0 & 0 & 0 & 0 & 0 & 0 & 0  \\
        1 & 1 & 1 & 0 & 1 & 0 & 0 & 0  \\
        0 & 0 & 0 & 0 & 0 & 0 & 0 & 0  \\
        0 & 0 & 0 & 0 & 0 & 0 & 0 & 0  \\
        0 & 0 & 0 & 0 & 0 & 0 & 0 & 0  \\
    \end{array} }\right]
\end{equation}

\begin{figure*}
\centering
\includegraphics[width=0.7\linewidth]{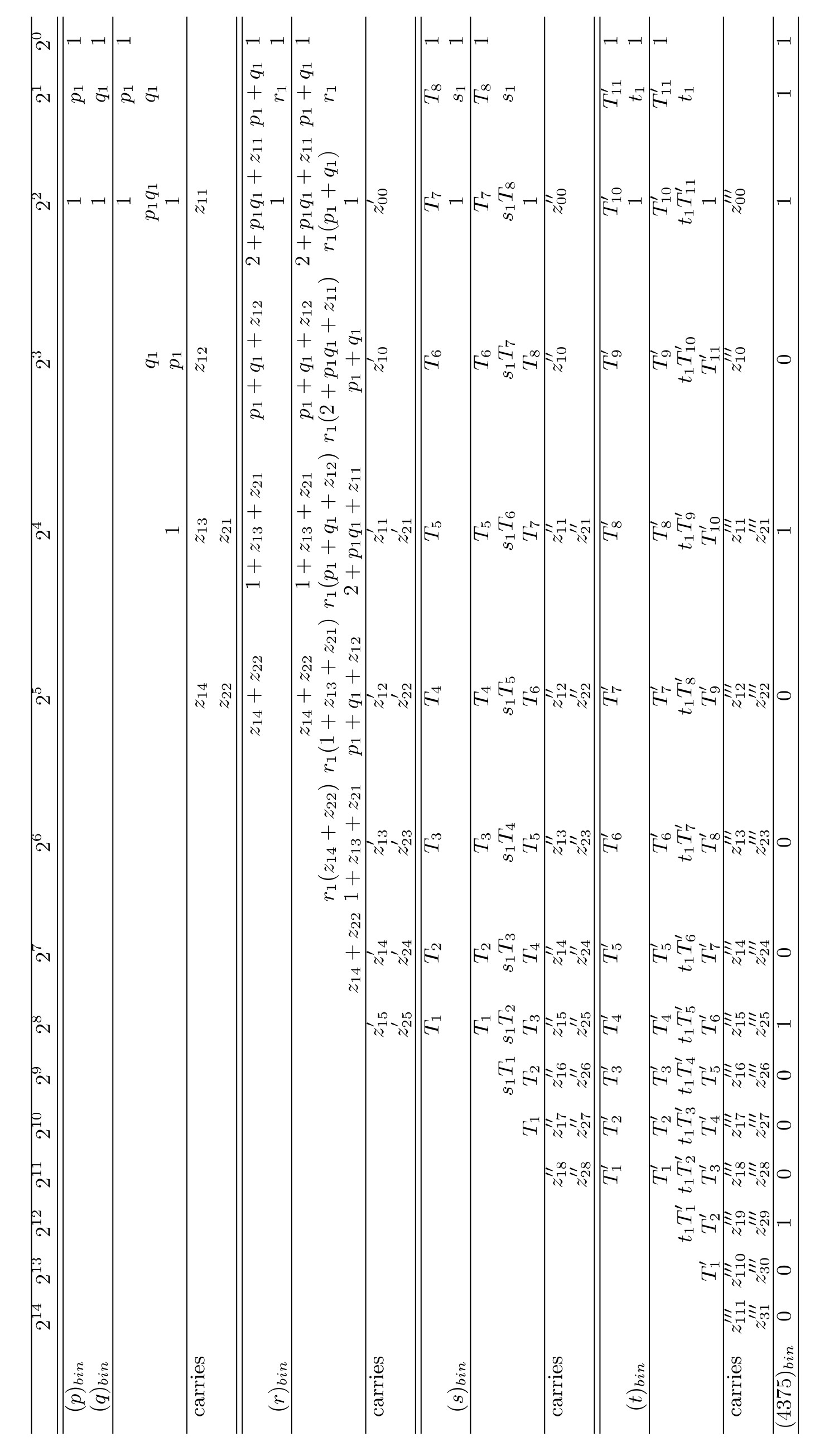} 
\caption{\emph{Multiplication table for $\alpha$=$5$ case, $N$=$4375$. The expressions for the variables are given in Eq. \ref{ex1}.}}
\label{t2}
\end{figure*}

\begin{figure*}
\centering
\includegraphics[width=\linewidth]{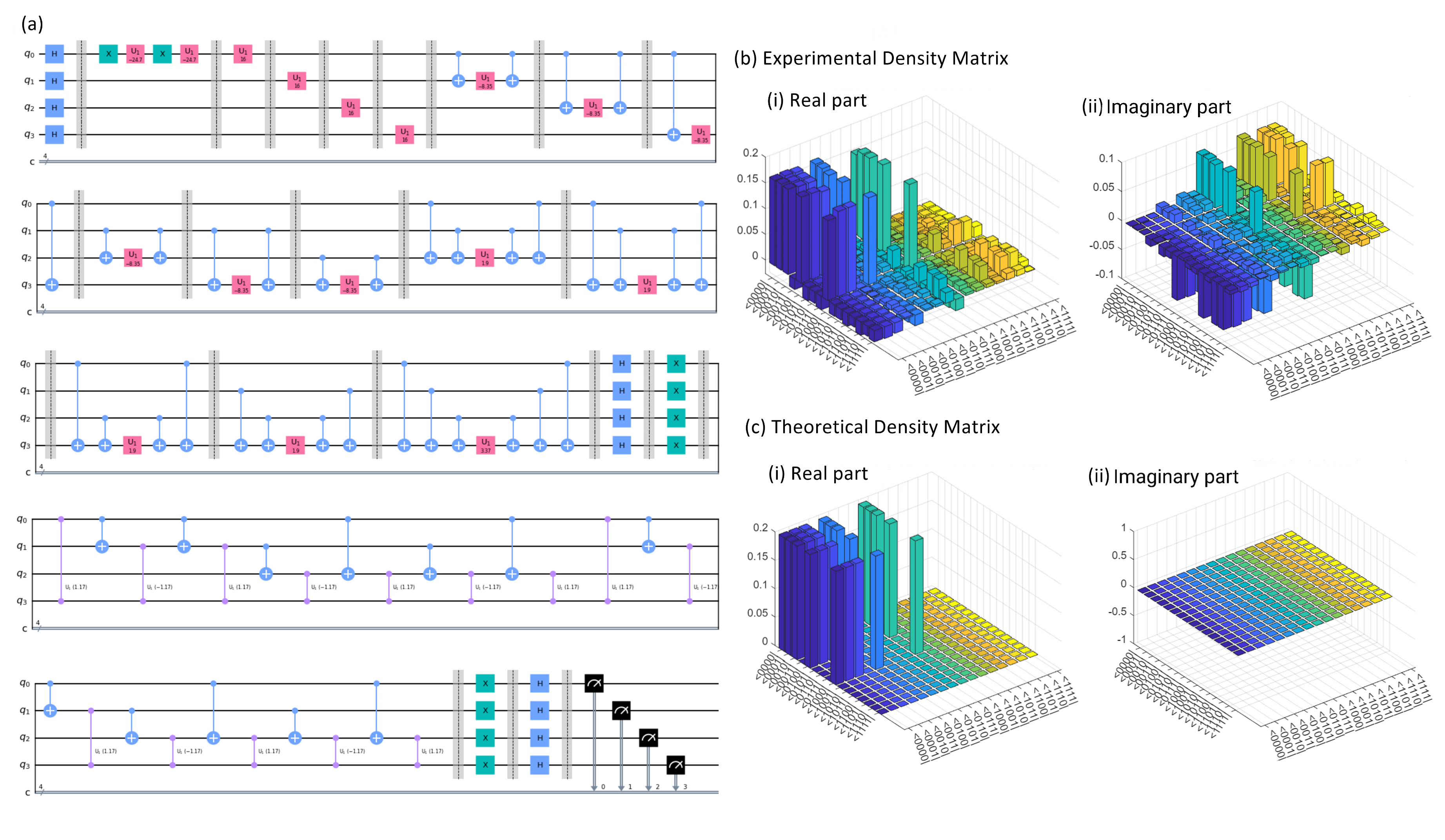} 
\caption{\textbf{(a)Quantum factorization circuit, (b) Experimental density matrix (EDM), and (c) Theoretical density matrix (TDM)} The circuit, which uses $4$ qubits to factorize $N$=$4375$, is implemented on IBM's quantum processor. The experiment and simulation are performed on the ibmq$\_$perth system, a $7$-qubit processor. The final measurements are obtained in the $Z$-basis, which gives the EDM. The EDM has a non-zero complex part, as shown in (b)(ii). As indicated in (b) and (c), the numerical values of the real component and imaginary component of the density matrix elements are shown separately. The fidelity of the result was found to be 0.9081. The expansion of $\hat{H}$ in Pauli basis only involves $I$ and $Z$ Pauli matrices. They all commute with one another since they are diagonal matrices. Therefore, the product of particular unitary gates can be used to express the exponential Hamiltonian function \cite{qfa_WhitfieldarXiv2010}. In order to get an equal superposition of states, $H^{\otimes n}$ is applied at the beginning. The first part of the circuit between $U$=$H^{\otimes n}$ applies the exponential operator $e^{-i\hat{H}\theta}$. The remaining portion of the circuit offers the $\ket{0^{\otimes n}}$ state an $e^{-i\theta}$ phase shift before delivering the final solution state.}
\label{qfa_Fig2}
\end{figure*}
 
The aforementioned conclusion is further generalized for tetra prime numbers of the type $N=p^3q$, where the other prime numbers $p$ and $q$ differ by one binary digit. The simplification for $N$=$1269636549803$ is provided. For $N$=$1269636549803$, there are four prime factors are denoted as $\{1p_9p_8...p_11\}_{bin}$, $\{1q_9q_8...q_11\}_{bin}$, $\{1r_9r_8...r_11\}_{bin}$, and $\{1s_9s_8...s_11\}_{bin}$ in binary system. From minimization, it is obtained that $p_i$=$q_i$=$r_i$=$s_i$=$0$ for $i=3,4,6,7,8,9$, and $p_i=q_i=r_i=s_i=1$ for $i=2,5$. The final set of equation for $N$=$1269636549803$ is same as Eq. \eqref{qfa_eq_1} for $p_1$, $q_1$, $r_1$, and $s_1$. Further implementation is the same as the tetra prime quantum factorization protocol. The solutions obtained from the ground states of the Hamiltonian result in the values of prime numbers to be $p$=$1061$, $q$=$1061$, $r$=$1061$, and $s$=$1063$, which are the required prime factors.

\subsection{Quantum factorization of penta prime number}

Applying the quantum factorization protocol to a number with five prime factors, $N$=$4375$. The binary form of the prime factors are expressed as $\{1p_11\}_{bin}$, $\{1q_11\}_{bin}$, $\{1r_11\}_{bin}$, $\{1s_11\}_{bin}$, and $\{1t_11\}_{bin}$ respectively. The binary multiplication table for $N$=$4375$ is provided in Fig. \ref{t2}. The value of $T_i$ ($i$ goes from $1$ to $8$) variables are the same as the ones given in Fig. \ref{t1}. The remaining expressions are given as:

\begin{eqnarray}\label{ex1}
    T_1'= z_{18}''+z_{28}''\nonumber \\ 
    T_2'= T_1+z_{17}''+z_{27}''\nonumber \\
    T_3'= s_1T_1+T_2+z_{16}''+z_{26}''\nonumber \\
    T_4'= T_1+s_1T_2+Tr_3+z_{15}''+z_{25}''\nonumber \\
    T_5'= T_2+s_1T_3+Tr_4+z_{14}''+z_{24}''\nonumber \\
    T_6'= T_3+s_1T_4+Tr_5+z_{13}''+z_{23}''\nonumber \\
    T_7'= T_4+s_1T_5+Tr_6+z_{12}''+z_{22}''\nonumber \\
    T_8'= T_5+sr_1T_6+T_7+z_{11}''+z_{21}''\nonumber \\
    T_9'= T_6+s_1Tr_7+T_8+z_{10}''\nonumber \\
    T_{10}'= T_7+s_1T_8+1+z_{00}''\nonumber \\
    T_{11}'= T_8+s_1
\end{eqnarray}

The set of equations found after applying the minimization procedure to the equations obtained from binary multiplication is given by:

 \begin{eqnarray}\label{e1}
    && p_1 + q_1 + r_1 + s_1 + t_1 = 1\nonumber \\
    && p_1q_1 + q_1r_1 + p_1r_1 + s_1p_1 + s_1q_1 + s_1r_1 + p_1t_1 + q_1t_1\nonumber \\
    &&+ s_1t_1 + r_1t_1 = 0 \nonumber \\
    && p_1q_1r_1 + p_1r_1s_1 + p_1q_1s_1 + q_1r_1s_1 + p_1q_1t_1 + p_1r_1t_1\nonumber \\
    &&+ q_1r_1t_1 + p_1s_1t_1 + q_1s_1t_1 + r_1s_1t_1 = 0 \nonumber \\
    && p_1q_1r_1s_1 + q_1r_1s_1t_1 + p_1r_1s_1t_1 + p_1q_1s_1t_1 + p_1q_1r_1t_1 = 0 \nonumber \\
\end{eqnarray}
 
 The set of equations is further simplified to a single equation in four variables:
 \begin{eqnarray}\label{qfa_eq_10}
    && p_1+q_1+r_1+s_1-3p_1^2-3q_1^2-3r_1^2-3s_1^2+3p_1^3+3q_1^3\nonumber \\ 
    &&+3r_1^3+3s_1^3-p_1^4-q_1^4-r_1^4-s_1^4-6p_1q_1-6r_1s_1-6p_1r_1\nonumber \\
    &&-6p_1s_1-6q_1r_1-6q_1s_1+9p_1^2q_1+9p_1^2r_1+9p_1^2s_1+9q_1^2p_1\nonumber \\
    &&+9q_1^2r_1+9q_1^2s_1+9r_1^2p_1+9r_1^2q_1+9r_1^2s_1+9s_1^2p_1+9s_1^2q_1\nonumber \\
    &&+9s_1^2r_1+18p_1q_1r_1+18p_1r_1s_1+18p_1q_1s_1+18q_1r_1s_1\nonumber \\
    &&-4p_1^3q_1-4p_1^3r_1-4p_1^3s_1-4q_1^3p_1-4q_1^3r_1-4q_1^3s_1\nonumber \\
    &&-4r_1^3p_1-4r_1^3q_1-4r_1^3s_1-4s_1^3p_1-4s_1^3q_1-4s_1^3r_1\nonumber \\
    &&-6p_1^2q_1^2-6p_1^2r_1^2-6p_1^2s_1^2-6q_1^2r_1^2-6q_1^2s_1^2\nonumber \\
    &&-6r_1^2s_1^2-12p_1q_1^2r_1-12p_1q_1^2s_1-12p_1r_1^2q_1-12p_1r_1^2s_1\nonumber \\
    &&-12p_1s_1^2q_1-12p_1s_1^2r_1-12q_1p_1^2r_1-12s_1p_1^2r_1\nonumber \\
    &&-12q_1p_1^2s_1-12q_1r_1^2s_1-12q_1s_1^2r_1-12s_1q_1^2r_1\nonumber \\
    &&-23p_1q_1r_1s_1=0
\end{eqnarray}

To form the Hamiltonian, $46p_1q_1r_1s_1$ is added to the left side of Eq. \ref{qfa_eq_10} and then the equation expression is multiplied by $-1$. The variables $(p_1,q_1,r_1,s_1)$ are mapped with the operators $(\hat{a_1},\hat{a_2},\hat{a_3},\hat{a_4})$ to encode the required states in the Hamiltonian:

 \begin{eqnarray}\label{qfa_eq_11}
    \hat{H} =&& -23(\hat{a_1}\hat{a_2}\hat{a_3}\hat{a_4})+18(\hat{a_1}\hat{a_2}\hat{a_3}+\hat{a_1}\hat{a_2}\hat{a_4}\nonumber \\
    &&+\hat{a_2}\hat{a_3}\hat{a_4}+\hat{a_1}\hat{a_3}\hat{a_4})+2(\hat{a_1}\hat{a_2}+\hat{a_1}\hat{a_3}+\hat{a_2}\hat{a_3}\nonumber \\
    &&+\hat{a_3}\hat{a_4}+\hat{a_1}\hat{a_4}+\hat{a_2}\hat{a_4})
\end{eqnarray}
 
The mapping encodes the solution in the ground states (states with zero eigenvalues) of the Hamiltonian of the factorization problem. Applying the definition of $\hat{a_i}$ operators, the Hamiltonian in Eq. \ref{qfa_eq_11} is given by:
 
\begin{eqnarray}\label{qfa_eq_12}
      \hat{H} =&& \frac{169}{16}(I_{4})-\frac{109}{16}(\sigma_z^1\otimes I\otimes I\otimes I+I\otimes\sigma_z^2\otimes I\otimes I\nonumber\\
      &&+I\otimes I\otimes \sigma_z^3\otimes I+I\otimes I\otimes I\otimes \sigma_z^4)\nonumber\\
      &&+\frac{57}{16}(\sigma_z^1\otimes\sigma_z^2\otimes I\otimes I+\sigma_z^1\otimes I\otimes \sigma_z^3\otimes I\nonumber\\
      &&+I\otimes\sigma_z^2\otimes \sigma_z^3\otimes I+\sigma_z^1\otimes I\otimes I\otimes \sigma_z^4\nonumber\\
      &&+I\otimes\sigma_z^2\otimes I\otimes \sigma_z^4+I\otimes I\otimes \sigma_z^3\otimes \sigma_z^4)\nonumber\\
      &&-\frac{13}{16}(\sigma_z^1\otimes \sigma_z^2\otimes \sigma_z^3\otimes I+\sigma_z^1\otimes \sigma_z^2\otimes I\otimes \sigma_z^4\nonumber\\
      &&+\sigma_z^1\otimes I\otimes \sigma_z^3\otimes \sigma_z^4+I\otimes \sigma_z^2\otimes \sigma_z^3\otimes\sigma_z^4)\nonumber\\
      &&-\frac{23}{16}(\sigma_z^1\otimes \sigma_z^2\otimes \sigma_z^3\otimes \sigma_z^4)
\end{eqnarray}

where $I_4$ is the $4$-qubit identity operator.
 
\begin{equation}\label{H2}
\tiny
    \hat{H}= \left[ {\begin{array}{cccc cccc cccc cccc}
        0 & 0 & 0 & 0 & 0 & 0 & 0 & 0 & 0 & 0 & 0 & 0 & 0 & 0 & 0 & 0   \\
        0 & 0 & 0 & 0 & 0 & 0 & 0 & 0 & 0 & 0 & 0 & 0 & 0 & 0 & 0 & 0    \\
        0 & 0 & 0 & 0 & 0 & 0 & 0 & 0 & 0 & 0 & 0 & 0 & 0 & 0 & 0 & 0    \\
        0 & 0 & 0 & 2 & 0 & 0 & 0 & 0 & 0 & 0 & 0 & 0 & 0 & 0 & 0 & 0   \\
        0 & 0 & 0 & 0 & 0 & 0 & 0 & 0 & 0 & 0 & 0 & 0 & 0 & 0 & 0 & 0   \\
        0 & 0 & 0 & 0 & 0 & 2 & 0 & 0 & 0 & 0 & 0 & 0 & 0 & 0 & 0 & 0    \\
        0 & 0 & 0 & 0 & 0 & 0 & 2 & 0 & 0 & 0 & 0 & 0 & 0 & 0 & 0 & 0   \\
        0 & 0 & 0 & 0 & 0 & 0 & 0 & 24 & 0 & 0 & 0 & 0 & 0 & 0 & 0 & 0  \\
        0 & 0 & 0 & 0 & 0 & 0 & 0 & 0 & 0 & 0 & 0 & 0 & 0 & 0 & 0 & 0  \\
        0 & 0 & 0 & 0 & 0 & 0 & 0 & 0 & 0 & 2 & 0 & 0 & 0 & 0 & 0 & 0   \\
        0 & 0 & 0 & 0 & 0 & 0 & 0 & 0 & 0 & 0 & 2 & 0 & 0 & 0 & 0 & 0   \\
        0 & 0 & 0 & 0 & 0 & 0 & 0 & 0 & 0 & 0 & 0 & 24 & 0 & 0 & 0 & 0  \\
        0 & 0 & 0 & 0 & 0 & 0 & 0 & 0 & 0 & 0 & 0 & 0 & 2 & 0 & 0 & 0   \\
        0 & 0 & 0 & 0 & 0 & 0 & 0 & 0 & 0 & 0 & 0 & 0 & 0 & 24 & 0 & 0  \\
        0 & 0 & 0 & 0 & 0 & 0 & 0 & 0 & 0 & 0 & 0 & 0 & 0 & 0 & 24 & 0  \\
        0 & 0 & 0 & 0 & 0 & 0 & 0 & 0 & 0 & 0 & 0 & 0 & 0 & 0 & 0 & 61   \\
    \end{array} }\right]
\end{equation}
 
\begin{table}
\centering
\caption{\emph{For $N$=$4375$. The phase shift relative to the state $\ket{0000}$ when the operator $e^{-i\hat{H}\theta}$ is applied to the $z$-basis states.}}
\begin{tabular}{cccc}
\hline
\hline
Quantum  & Relative & Quantum & Relative \\
State  & Phase shift & State & Phase shift \\ \hline
$\ket{0000}$   & $0$         & $\ket{1000}$   & $0$   \\
$\ket{0001}$   & $0$         & $\ket{1001}$   & $-2\theta$   \\ 
$\ket{0010}$   & $0$         & $\ket{1010}$   & $-2\theta$   \\
$\ket{0011}$   & $-2\theta$  & $\ket{1011}$   & $-24\theta$   \\
$\ket{0100}$   & $0$         & $\ket{1100}$   & $-2\theta$   \\
$\ket{0101}$   & $-2\theta$  & $\ket{1101}$   & $-24\theta$   \\ 
$\ket{0110}$   & $-2\theta$  & $\ket{1110}$   & $-24\theta$   \\
$\ket{0111}$   & $-24\theta$  & $\ket{1111}$   & $-61\theta$   \\
\hline
\hline
\end{tabular}
\label{qft2}
\end{table}
 
The ground states of the above Hamiltonian (Eq. \ref{H2}) give the required bit solutions. The ground states ($|p_1q_1r_1s_1\rangle$) are $\ket{0000}$, $\ket{0001}$, $\ket{0010}$, $\ket{0100}$, and $\ket{1000}$, which gives factors $5$, $5$, $5$, $5$, and $7$ after inputting the values of $p_1$, $q_1$, $r_1$, and $s_1$ in Eq. \ref{e1}. The phase shift angle is specified as $\theta=2\sin^{-1}{\left(\frac{\sin{\frac{\pi}{4j+2}}}{\sin{\phi}}\right)}$ where $\phi$=$\sin^{-1}{\left(\frac{\sqrt{5}}{4}\right)}$ and $j$=$2$ iterations for the search algorithm to function and amplify the desired results \cite{qfa_LiuIJTP2014,qfa_liPLA2007}. The EDM is given in Fig. \ref{qfa_Fig2} (b).  The TDM is given by $\rho^T$=$\ket{\Psi}\bra{\Psi}$, where $\ket{\Psi}$=$\frac{1}{\sqrt{5}}\left[\ket{0000}+\ket{0001}+\ket{0010}+\ket{0100}+\ket{1000}\right]$.
 \begin{equation}\label{qfa_eq_13}
    \rho^{T}= \frac{1}{5} \left[ {\begin{array}{ccccccccccccc}
        1 & 1 & 1 & 0 & 1 & 0 & 0 & 0 & 1 & 0 & . & . & 0  \\
        1 & 1 & 1 & 0 & 1 & 0 & 0 & 0 & 1 & 0 & . & . & 0  \\
        1 & 1 & 1 & 0 & 1 & 0 & 0 & 0 & 1 & 0 & . & . & 0  \\
        0 & 0 & 0 & 0 & 0 & 0 & 0 & 0 & 0 & 0 & . & . & 0   \\
        1 & 1 & 1 & 0 & 1 & 0 & 0 & 0 & 1 & 0 & . & . & 0  \\
        0 & 0 & 0 & 0 & 0 & 0 & 0 & 0 & 0 & 0 & . & . & 0   \\
        0 & 0 & 0 & 0 & 0 & 0 & 0 & 0 & 0 & 0 & . & . & 0   \\
        0 & 0 & 0 & 0 & 0 & 0 & 0 & 0 & 0 & 0 & . & . & 0   \\
        1 & 1 & 1 & 0 & 1 & 0 & 0 & 0 & 1 & 0 & . & . & 0  \\
        0 & 0 & 0 & 0 & 0 & 0 & 0 & 0 & 0 & 0 & . & . & 0   \\
        0 & 0 & 0 & 0 & 0 & 0 & 0 & 0 & 0 & 0 & . & . & 0   \\
        0 & 0 & 0 & 0 & 0 & 0 & 0 & 0 & 0 & 0 & . & . & 0   \\
        0 & 0 & 0 & 0 & 0 & 0 & 0 & 0 & 0 & 0 & . & . & 0   \\
        0 & 0 & 0 & 0 & 0 & 0 & 0 & 0 & 0 & 0 & . & . & 0   \\
        0 & 0 & 0 & 0 & 0 & 0 & 0 & 0 & 0 & 0 & . & . & 0   \\
        0 & 0 & 0 & 0 & 0 & 0 & 0 & 0 & 0 & 0 & . & . & 0   \\
    \end{array} }\right]
\end{equation}
 
The quantum factorization of tetra and penta prime numbers provides results with high fidelity and can therefore be extended to larger integers. A general set of equations for the minimization part is provided for larger numbers that share similar properties in their prime factors. For $N$=$({n_1})_{bin} \times ({n_2})_{bin}…({n_{\alpha}})_{bin}$. Let $({n_j})_{1}$, where $j$ going from $1$ to $\alpha$, and $1$ denote the binary digit position. The binary position subscript ($1$) is dropped to avoid confusion. The form of the set of equations is shown in Table \ref{qtf} using the property mentioned regarding the number of digits in the binary number and Eq. \ref{equn1}. The final expression obtained for each case might need some modification to achieve the proper expression for the Hamiltonian of the factorization problem.

\begin{table}
\centering
\caption{\emph{The equations' values for the two scenarios are given. In instances $1$ and $2$, $n_1$=$1$, $n_2$=$n_3$=...=$n_{\alpha}$=$0$ and $n_1$=$0$, $n_2$=$n_3$=...=$n_{\alpha}$=$1$ respectively.}}
\begin{tabular}{ccc}
\hline
\hline
Equations  & Case $1$ & Case $2$  \\ \hline
${n_1}+{n_2}+...+{n_{\alpha}}$   & $1$         & $\alpha-1$    \\
$\Sigma_{i_1<i_2}{n_{i_1}}{n_{i_2}}$   & $0$         & $\binom{\alpha}{2}-\binom{\alpha-1}{1}$    \\ 
$\Sigma_{i_1<i_2<i_3}{n_{i_1}}{n_{i_2}}{n_{i_3}}$   & $0$ & $\binom{\alpha}{3}-\binom{\alpha-1}{2}$   \\
. & . & .   \\
. & . & .   \\
. & . & .   \\
$\Sigma_{i_1<i_2<...<i_{\alpha-1}}{n_{i_1}}{n_{i_2}}....{n_{i_{\alpha-1}}}$   & $0$ & $\binom{\alpha}{\alpha-1}-\binom{\alpha-1}{\alpha-2}$   \\
\hline
\hline
\end{tabular}
\label{qtf}
\end{table}

\section{Discussion and Conclusion}

Many factorization problems were solved using different methods, including quantum annealing properties and the adiabatic principles \cite{qfa_WangFP2022,qfa_DattaniarXiv2014,qfa_DridiScirep2017,qfa_XuPRL2012,qfa_LiarXiv2017,qfa_DasharXiv2018}. The exponential function of the Hamiltonian operator was used in this article to mark the Hamiltonian's ground states as opposed to the quantum annealing technique for the factorization problem. Following minimization, the final equation is used to create the Hamiltonian operator. To amplify the marked states, the generalized Grover's algorithm was implemented \cite{qfa_LiuIJTP2014}. For the proof-of-concept, these numbers $875$, $1269636549803$, and $4375$ were factorized where the first two numbers have four prime numbers, and the last number has five prime numbers that were factorized using the $7$-qubit IBM quantum processor (Perth). The processor type of Perth is Falcon r5.11H. The Falcon family of devices is proven to be advantageous for medium-scale circuits. H represents the segment consisting of the chip-sub sections and is defined differently for each processor family. The r5.11 is the most updated version of the processor in the Falcon family. For $N$=$875$, the ground states of the Hamiltonian were found to be $\ket{000}$, $\ket{001}$, $\ket{010}$, and $\ket{100}$ that corresponded to the prime numbers $5$, $5$, $5$, and $7$. The fidelity of the quantum circuit for $N$=$875$ was estimated to be 0.9683. For $N$=$1269636549803$, the set of equations from the binary product has more variables, but the final equation obtained after minimization is the same. Hence, the quantum circuit for $N$=$1269636549803$ is the same as that used for $N$=$875$. For $N$=$4375$, the ground states of the Hamiltonian were found to be $\ket{0000}$, $\ket{0001}$, $\ket{0010}$, $\ket{0100}$, and $\ket{1000}$ that corresponded to the prime numbers $5$, $5$, $5$, $5$, and $7$. The fidelity of the quantum circuit for $N$=$4375$ was calculated to be 0.9081. The two quantum circuits' high fidelity assures the factorization protocol's feasibility with large prime factors and more prime factors. These findings are particularly significant because the online security system is predicated on the hypothesis that factoring big numbers is an NP-hard task. Adding further to the interest in studying and implementing quantum computation techniques to build more secure systems. The optimized set of equations for larger numbers is provided at the end, bearing the same property of prime factors as the number factorized in this article. The number of qubits to solve a factorization problem depends on the simplification. The pre-processing part of the factorization problem (performing the binary product of binary numbers) is done with the help of a computer program. The other quantum factorization methods require either a large number of qubits or many iterations for evolution for implementation, but that isn't the case for the protocol used in this article. This is the first experimental realization of quantum algorithms to factor a number with four and five prime factors.




\end{document}